\documentclass[aps,superscriptaddress,showpacs]{revtex4}
\usepackage{epsfig}
\usepackage{graphicx}
\usepackage{amsmath}
\RequirePackage{xspace}

\def\BR         {{\ensuremath{\cal B}\xspace}}

\begin{document}

\title{\bf Study of the $e^+e^-\to\eta\gamma$ process with SND detector  
\\ at the VEPP-2M $e^+e^-$ collider}
\author{M.~N.~Achasov}
\author{V.~M.~Aulchenko}
\author{K.~I.~Beloborodov}
\affiliation{Budker Institute of Nuclear Physics,  Novosibirsk 630090, Russia}
\affiliation{Novosibirsk State University, Novosibirsk 630090, Russia}
\author{A.~V.~Berdyugin}
\email{berdugin@inp.nsk.su}
\author{A.~G.~Bogdanchikov}
\author{A.~V.~Bozhenok}
\author{D.~A.~Bukin}
\author{T.~V.~Dimova}
\affiliation{Budker Institute of Nuclear Physics,  Novosibirsk 630090, Russia}
\author{V.~P.~Druzhinin}
\author{V.~B.~Golubev}
\affiliation{Budker Institute of Nuclear Physics,  Novosibirsk 630090, Russia}
\affiliation{Novosibirsk State University, Novosibirsk 630090, Russia}
\author{A.~A.~Korol}
\author{S.~V.~Koshuba}
\author{A.~V.~Otboev}
\author{E.~V.~Pakhtusova}
\affiliation{Budker Institute of Nuclear Physics,  Novosibirsk 630090, Russia}
\author{S.~I.~Serednyakov}
\author{Yu.~M.~Shatunov}
\affiliation{Budker Institute of Nuclear Physics,  Novosibirsk 630090, Russia}
\affiliation{Novosibirsk State University, Novosibirsk 630090, Russia}
\author{V.~A.~Sidorov}
\affiliation{Budker Institute of Nuclear Physics,  Novosibirsk 630090, Russia}
\author{Z.~K.~Silagadze}
\affiliation{Budker Institute of Nuclear Physics,  Novosibirsk 630090, Russia}
\affiliation{Novosibirsk State University, Novosibirsk 630090, Russia}
\author{A.~N.~Skrinsky} 
\affiliation{Budker Institute of Nuclear Physics,  Novosibirsk 630090, Russia}
\author{A.~V.~Vasiljev}
\affiliation{Budker Institute of Nuclear Physics,  Novosibirsk 630090, Russia}
\affiliation{Novosibirsk State University, Novosibirsk 630090, Russia}

\begin{abstract}
In experiment with the SND detector at  the VEPP-2M $e^+e^-$ collider 
the $e^+e^-\to\eta\gamma$ cross section was measured in the center-of-mass 
energy range $E$=0.60--1.38 GeV with the integrated luminosity of 
27.8~pb$^{-1}$. The measured cross section is well described by the vector
meson dominance model with contributions from the $\rho(770)$, 
$\omega(783)$, $\phi(1020)$, $\rho^{\prime}(1465)$ resonances and agrees 
with results of previous measurements. The decay probabilities 
$\BR(\phi\to\eta\gamma)$, $\BR(\omega\to\eta\gamma)$ and $\BR(\rho\to\eta\gamma)$
were measured with the accuracies better than or comparable to the 
world averages.
\end{abstract}


\maketitle

\section{Introduction}
Measurements of the vector mesons radiative decays 
$\rho,~\omega,~\phi\to\pi^0\gamma,~\eta\gamma$ were subject of 
experimental investigation in several tens of experiments 
during more than 40 years \cite{pdg}. However, further improvement 
of accuracy of these branching ratios measurements is still 
important for development of various phenomenological models --- 
quark models, SU(3) based linear and non-linear sigma models and vector 
meson dominance models, \cite {link1, link2, link3}.

The vector meson radiative decays give information on the underlying 
non-perturbative QCD dynamics and this is one reason why interest in such 
decays is still not exhausted. The $\eta$ meson is a member of the low-lying
pseudoscalar octet and therefore is a would-be Goldstone boson associated
with the spontaneous breaking of chiral symmetry in the QCD vacuum. Besides
it is significantly connected to the $\eta^\prime$ meson which by itself
is strongly affected by QCD axial anomaly. Therefore any information about
the $\eta$ meson structure, and the vector mesons radiative decays are one of
sources of such information, will give a clue about the QCD vacuum and 
mechanisms of the chiral symmetry breaking.

Theoretically radiative decays with the $\eta\gamma$ final state were investigated
in the context of a quark-level linear sigma model \cite{QCD1}, using QCD sum 
rules \cite{QCD2} and in the framework of the non-relativistic quark model 
\cite{QCD3}.

Another interesting theoretical problem where the high precision experimental
input from the vector meson radiative decays is welcome is the 
$\eta-\eta^\prime$ mixing problem. As the experimental data became more 
precise it turned out that the traditional one mixing angle scheme does not
work properly and more sophisticated two mixing angle description was 
developed \cite{mixing1}. The physics underlying $\eta$ and $\eta^\prime$ 
mesons constitutes a vivid and fascinating research field today providing
unexpected challenges and surprises \cite{mixing2}. 

The best accuracy in measurement of the decay probabilities $\rho,~\omega,
~\phi\to\eta\gamma$ was achieved in the last $e^+e^-$ storage ring 
experiments with CMD-2 \cite{exd3,exd2,exd23} and SND 
\cite{exd16,exd7,exd17,exd22,exd18} detectors through investigation 
of the $e^+e^-\to\rho,~\omega,~\phi\to\eta\gamma$ processes.
The reached accuracy is of the order of 10\% for the decays $\rho,~\omega 
\to\eta\gamma$ and about 2\% for the $\phi\to\eta\gamma$ decay, and the 
last result has been obtained by averaging more than 10 measurements with 
accuracies of the order of 5\% each.

In this paper we present results of our studies of the process 
\begin {eqnarray} 
e^+e^-\to\eta\gamma, \label {etg} 
\end {eqnarray} 
with the subsequent decays of the $\eta$ meson into the three-pion final 
states:
\begin {eqnarray} 
\eta \to 3 \pi^0 \label {etgnc} \\ 
\eta \to \pi^+ \pi^-\pi^0. \label {etgcc} 
\end {eqnarray} 
Experimental data with integrated luminosity of 27.8~ pb$^{-1}$ collected
in experiments with the SND detector at the VEPP-2M collider in the 
center-of-mass energy
range $E$=0.60--1.38 GeV were analyzed.

The aim of the present work is to increase the accuracy of measurements 
of the decay probabilities $\rho,~\omega,~\phi\to\eta\gamma$, and also 
to measure the cross section in the non-resonant region, in particular, at 
energies above the $\phi$-meson resonance.

\section{The SND detector}
The general purpose non-magnetic detector SND \cite {SND} was developed 
for experiments at VEPP-2M $e^+e^-$ collider. The basic part of SND 
is a three layer electromagnetic calorimeter consisting of 1632 NaI(Tl)
crystals. Total thickness of the calorimeter is 13.4 radiation 
lengths. The calorimeter covers nearly $90\%$ of the full solid angle: 
$18^\circ\le\theta\le 162^\circ $, where $\theta $ is the polar angle.
Dependence of the energy resolution of the calorimeter on energy of the 
photon is given by the formula 
$\sigma_E/E _{\gamma}(\%)=4.2\%/\sqrt[4]{E_{\gamma}(\mbox{GeV})}$, 
while the angular resolution is 
$\sigma_{\varphi}=0.82/\sqrt{E _{\gamma}(\mbox {GeV})}\oplus 
0.63$ degrees.

For determination of charged particles production angles, two coaxial 
cylindrical drift chambers are used. The angular resolution is $1.8^\circ$ 
in polar direction, and $0.53^\circ$ in the azimuthal direction.

Experiments were performed in 1995--2000 in the energy range $E$=0.38-- 1.38~GeV.
The statistics was collected by repeated scanning 
of the energy range with variable step. Integrated luminosity was measured 
using elastic scattering and two-photon annihilation with 
2\% accuracy. In total the SND detector recorded about $1.5\times 10^9$ 
events. From them about $10^7$ are events with $\phi$-meson production
and decay, $3.7\times 10^6$ with the 
$\omega$-meson production, and $7\times 10^6$ with the $\rho$-meson
production.

\section{Selection of events in the decay channel $\eta\to 3\pi^0$ 
\label {evsel}}
\begin{figure}
\includegraphics[width=0.9\linewidth]{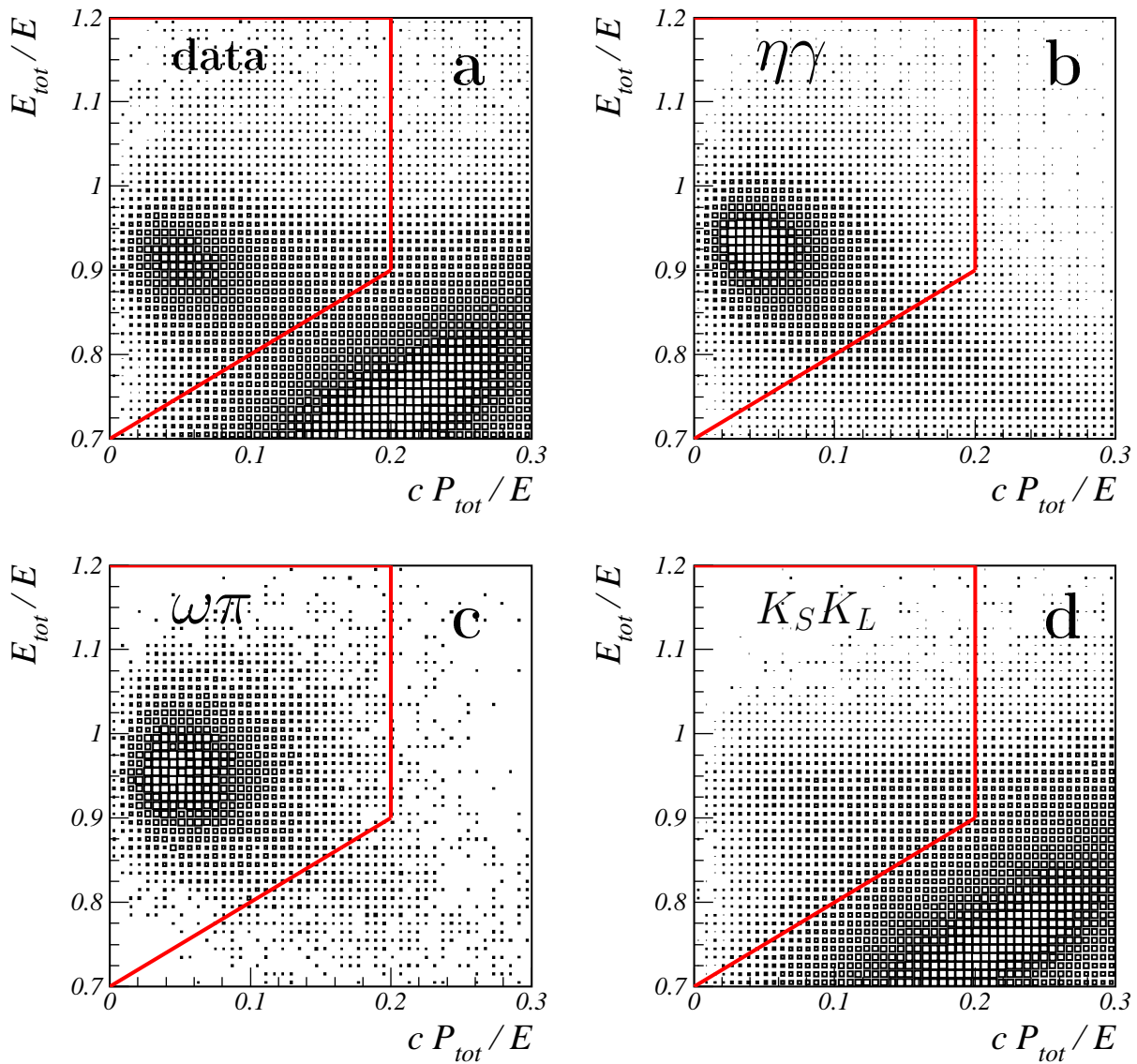}
\caption{ \label{eton_ptrt}
Two-dimensional distributions of normalized calorimeter energy deposition vs.
normalized total momentum of the detected particles for the selected data 
events in the whole energy region (a) and for simulated events of the signal and background 
processes (b,c,d). Solid lines indicate the selection cuts.
}
\end{figure}

Selection of events of the process under study (\ref{etg}) in the decay
mode (\ref{etgnc}) was performed in several steps. At the first stage 
the events satisfying the following conditions were selected:
\begin{enumerate}
\item $N_{\gamma} \geq 6$ , where $N_{\gamma}$ is the number of 
reconstructed photons in an event;
\item $N_{c} = 0$ , where $N_{c}$ is the number of reconstructed charged 
particle tracks;
\item $0.7 < E_{tot}/E < 1.2$ , where $E_{tot}$ is the total energy 
deposition in the calorimeter;
\item $cP_{tot}/E < 0.2$ , where $P_{tot}$ is the total momentum of photons;
\item $(E_{tot}-cP_{tot})/E > 0.7$.
\end{enumerate}
We allow a loss of one soft photon from 7 final photons of the
$e^+e^-\to\eta\gamma\to 3\pi^0\gamma$ reaction. Besides, we accept events with 
extra photons ($N _ {\gamma}>7$). The extra photons 
appear as a result of shower splitting in the calorimeter
or superimposed machine background. Since the final state of the process 
under study includes only photons, the energy deposition in the calorimeter 
and the total photon momentum (magnitude of the vector sum of photon momenta)
are close to $e^+e^-$ center-of-mass energy and zero, respectively.
Fig.\ref {eton_ptrt} shows two-dimensional distribution of these parameters
for data events, and Monte-Carlo (MC) simulated events of signal and background processes.
The line indicates the selection cuts 3, 4 and 5. 
With the use of these criteria, 47676 events have been selected.

As the background sources one has the following processes 
\begin {eqnarray} e^+ e^-\to K_S K_L, \; 
K_S \to 2 \pi^0 \label {ksl}, \\ 
e^+e ^-\to\pi^0\pi^0\gamma, \label {ppg} \\ 
e^+e^-\to\omega\pi^0\pi^0, \; \omega\to\pi^0 \gamma. \label {opp} 
\end {eqnarray} 
Analysis  of the experimental data has shown that the quantum-electrodynamic 
events from processes with large cross sections, for example:
\begin {eqnarray} e^+e^-\to 2(3)\gamma, \label {ggg} 
\end {eqnarray} 
also can give the required event configuration when the machine background
is superimposed on them. The contributions of other background processes are 
negligible.

For events selected at the first stage, kinematic fits 
using the measured photon angles and energies,  
and energy-momentum conservation laws, were performed. The quality of each
kinematic fit is characterized by the parameter $\chi^2$. 
Five kinematic hypotheses were checked:
\begin{enumerate}
\item $e^+e^-\to n\gamma,\; n\geq 6$ ($\chi^2_{n\gamma}$ -- the 
corresponding parameter),
\item $e^+e^-\to\eta\gamma\to 3\pi^0\gamma\to 7\gamma$ ($\chi^2_
{\eta\gamma}$),
\item $e^+e^-\to 3\gamma$ ($\chi^2_{3\gamma}$),
\item $e^+e^-\to\pi^0\pi^0\gamma$ ($\chi^2_{\pi^0\pi^0\gamma}$).
\item $e^+e^-\to\omega\pi^0\pi^0\to 3\pi^0\gamma\to 7\gamma$ 
($\chi^2_{\omega\pi^0\pi^0}$).
\end{enumerate}
In hypotheses 2, 4, 5 additional constraints on invariant masses of 
the photon pairs and on the invariant mass of the $\pi^0\gamma$ system 
were applied. In case when the number of detected photons in the event 
exceeded the number of photons necessary for the given hypothesis, kinematic 
fit was performed for all possible photon combinations and 
the one with the lowest $\chi^2$ value was selected.

In the energy region below 1~GeV the process $e^+e^-\to\eta\gamma,\,\eta\to
3 \pi^0$ is the only process with a significant cross section and 
multi-photon final state. In the $\phi$-meson region there is a contribution 
from the process (\ref{ksl}).
Since this energy region is above the $\omega\pi^0$ 
production threshold, there is also a background from 
the process (\ref{ppg}). 

\begin{figure}
\includegraphics[width=0.6\linewidth]{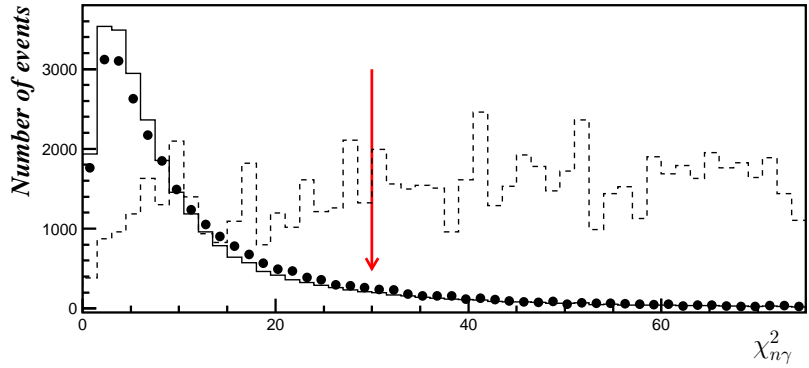}
\caption{\label{fig:chi} The $\chi^2_{n\gamma}$ distributions
for data events with $E<1.06$ GeV (points with error bars).
The dashed histogram shows the expected distribution for
background processes (\ref {ksl}) and (\ref {ppg}) scaled by a factor of 100.
The solid histogram shows the sum of simulated signal and background
distributions. The data and simulated distributions  are normalized to 
the sane area.
The cut on $\chi^2_{n\gamma}$ is indicated by the arrow.}
\end{figure}

For separation of events of the process (\ref {etg}) at $E<1.06$ GeV and 
suppression of the background from processes (\ref {ksl}), (\ref {ppg}) and 
(\ref {ggg}), the following cuts on the $\chi^2$ parameters 
were imposed:
\begin{enumerate}
\item $\chi^2_{n\gamma} < 30$ ;
\item $\chi^2_{3\gamma} > 20$ ;
\item $\chi^2_{\pi^0\pi^0\gamma} > 20$ .
\end{enumerate}
The distributions of $\chi^2_{n\gamma}$ for
data events and simulated signal and background events are shown in
Fig.~\ref{fig:chi}.

\begin{figure}
\includegraphics[width=0.6\linewidth]{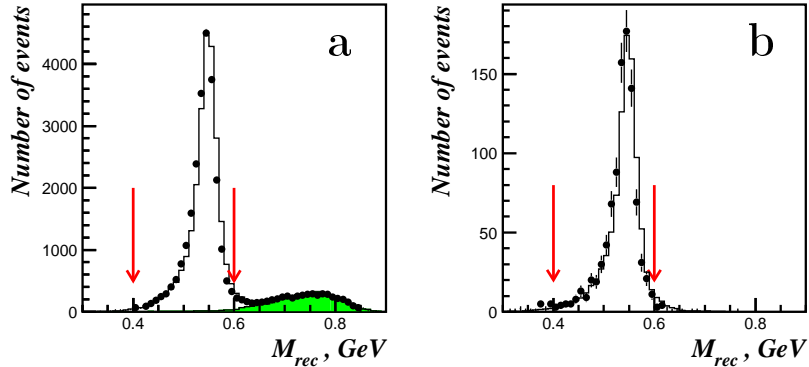}
\caption{\label{fig:ere1} Recoil mass of the most energetic photon in 
$e^+e^-\to n\gamma (n\geq 6)$ events. a) $\phi$-meson region; 
b) $\omega$ and $\rho$ mesons region. The histogram shows MC simulation of 
the process $e^+e^-\to\eta\gamma,~\eta\to 3\pi^0$. The shaded histogram shows 
simulation of all background processes. Points with error bars are data. 
Vertical arrows indicate the cut on 
$M_{rec}$ used for final selection of signal events.}
\end{figure}

After applying these selections criteria about $32\times10^3$ events
were left for further analysis. 
The recoil mass distribution of the most energetic photon 
$M_{rec}$ is presented in Fig.~\ref{fig:ere1}
for energies close to the $\omega$ and $\phi$ resonances. Events 
in Fig.\ref {fig:ere1}b with $M_{rec}>0.6$ GeV are attributed to
the  process (\ref {ksl}). Final selection of events was carried out 
with the cut $0.4<M_{rec}<0.6$ GeV.

The contribution of the background process (\ref {ksl}) for each energy 
point in the $\phi$-meson region was determined as follows:
\begin{equation}
N_{K_SK_L} = N_{K_SK_L}^{data}(0.6<M_{rec}<0.8)~\frac{N_{K_SK_L}^{MC}
(0.4<M_{rec}<0.6)}
{N_{K_SK_L}^{MC}(0.6<M_{rec}<0.8)}, \label{nkskl}
\end{equation}
where $N_{K_SK_L}^{MC}(0.4<M_{rec}<0.6)$ and $N_{K_SK_L}^{MC}
(0.6<M_{rec}<0.8)$ are
number of MC simulated events of the process (\ref {ksl}) with recoil mass
of the most energetic photon in intervals 0.4--0.6 GeV and 
0.6--0.8 GeV, respectively. $N_{K_SK_L}^{data}(0.6<M_{rec}<0.8) $ is the 
number of data events with $0.6<M_{rec}<0.8$ GeV.
The contribution of the background process (\ref {ksl}) for the energy 
points in the range 1.00 -- 1.03 GeV is about 1.2\% of the total number
of selected events at each point. The systematic error on 
the background event number from the process (\ref {ksl}) is 40\%. 
It was estimated as a difference between the calculation with the formula 
(\ref {nkskl}) and direct estimation by simulation.

In the energy range from 1.06 GeV (the $\omega\pi^0\pi^0 $ production  
threshold) up to 1.40 GeV the contribution of the background process 
(\ref {opp}) is comparable to that of signal process 
(\ref {etg}). The kinematics of events (\ref {opp}) 
closely resembles the kinematics of events (\ref {etg}), which complicates 
the suppression of the background related to the process (\ref {opp}).
Therefore, additional selection criteria were applied to events in the 
energy region $E>1.06$ GeV:
\begin{enumerate}
\item $N_{\gamma} = 7$,
\item $\chi^2_{\eta\gamma} < 60$,
\item $\chi^2_{\eta\gamma} - \chi^2_{\omega\pi^0\pi^0} < 0$ .
\end{enumerate}
With these conditions 47 events with energy $E>1.06$ GeV were selected
(Tables \ref{tabcrs1}, \ref{tabcrs2}).

The estimation based on measured cross section of the process 
(\ref {opp}) \cite {oppr} and its detection efficiency 
determined from the MC simulation for selection criteria described above
shows that the background from the process (\ref {opp}) is
negligible.

\section{Selection of events in the $\eta\to\pi^+\pi^-\pi^0$ decay mode}
For selection of events of the signal process (\ref{etg}) in the decay 
mode (\ref {etgcc}), the following selection criteria were applied at 
the first stage: 
\begin {enumerate} \item $N_c=2$,
\item $N_ {\gamma}\geq 3$, 
\item $R_i<0.25$cm, where $R_i$ is the distance between the $i$-th charged 
particle track and the beam axis,
\item $|Z_i| <10 $cm, where $Z_i$ is the coordinate of the charged particle
production point along the beam axis.
\end{enumerate}

The main background process at energies near and below the 
$\phi$ resonance is the process
\begin{eqnarray}
e^+e^-\to\pi^+\pi^-\pi^0 \label{pi3}
\end{eqnarray}
with an extra photon from
beam background, photon radiation by initial or final particles 
or nuclear interaction of the charged pions with the detector material.
In the energy region above the $\phi$ meson, the main background contribution
is expected from the process
\begin{eqnarray}
e^+e^-\to\pi^+\pi^-\pi^0\pi^0~. \label{pi4}
\end{eqnarray}
For energies above 1.06~GeV this background significantly 
exceeds the signal, thus the process (\ref {etg}) in the mode 
(\ref {etgcc}) was not studied at energies above 1.06~GeV.

For about $3.7\times10^5$ events selected with the above-stated criteria in the
energy range $E\leq 1.06$GeV, the kinematic fit was performed in two 
hypotheses:
\begin{enumerate}
\item $e^+e^-\to\pi^+\pi^-\pi^0\gamma\to\pi^+\pi^-\gamma\gamma\gamma$ 
($\chi^2_{3\pi\gamma}$),
\item $e^+e^-\to\pi^+\pi^-\pi^0\to\pi^+\pi^-\gamma\gamma$ ($\chi^2_{3\pi}$).
\end{enumerate}
For selection of events of the process $e^+e^-\to\pi^+\pi^-\pi^0\gamma$ 
and suppression of the background from the process (\ref {pi3}) 
the following cuts were used:
\begin{enumerate}
\item $\chi^2_{3\pi\gamma} < 25$,
\item $\chi^2_{3\pi} > 50$.
\end{enumerate}

\begin{figure}
\includegraphics[width=0.9\linewidth]{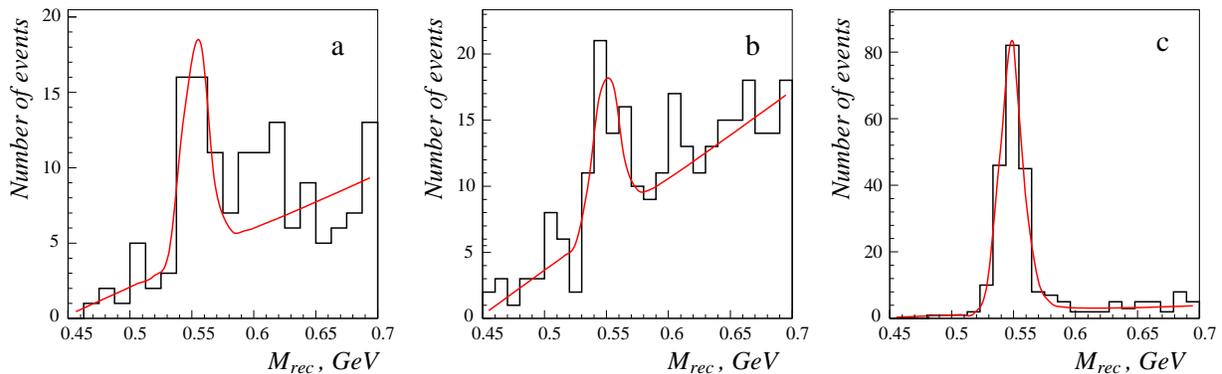}
\caption{ \label{gm3q}
Distributions of the recoil mass of the photon
not included into the reconstructed $\pi^0$-meson
for selected data events (histogram). The curve represents  
the result of the fit described in the text.
a) $E$ = 0.764--0.774 GeV;
b) $E$ = 0.784 GeV;
c) $E$ = 1.022 GeV.
}
\end{figure}

Distributions of the selected events over the recoil mass of the photon,
which is not included into the reconstructed $\pi^0$-meson, are shown in 
Fig.~\ref {gm3q} for three energy points. Events of the process (\ref {etg})
form a narrow peak near the $\eta$-meson mass. Events of the background 
process (\ref{pi3}) have the distribution which 
is well described by a linear function
in the recoil mass interval 0.45--0.7~GeV.

The number of events of the process $e^+e^-\to\eta\gamma\to\pi^+\pi^-\pi^0
\gamma$ in each energy point was determined by approximating the recoil mass
spectra by a sum of the signal and background distributions. For the 
description of the shape of the signal spectrum so called Novosibirsk
function \cite{nfun} was used which parameters (the width and asymmetry) were 
taken from simulation. Comparison of the $\eta$ line-shapes 
for the simulated and data events was 
performed in the energy range close to the $\phi$-meson resonance 
(Fig.~\ref {gm3q}c) and it was shown that the simulation 
reproduces the data shape with sufficient accuracy. The recoil photon spectrum 
for the background events was approximated either by a linear function or by 
a parabola. The difference between the results of approximations 
with two different shapes of the background was used for the estimation of 
the associated systematic error. The obtained numbers of events 
for different energy points are listed  in the Table~\ref{tabcrs3}. The quoted errors are
statistical and systematic, respectively. For a part of the data points 
the statistics of neighboring energy points were summed together. In this case 
the boundaries of the energy interval are shown in the Table ~\ref{tabcrs3}.
	  	
\section{Detection efficiency}
\begin{figure}
\includegraphics[width=0.9\linewidth]{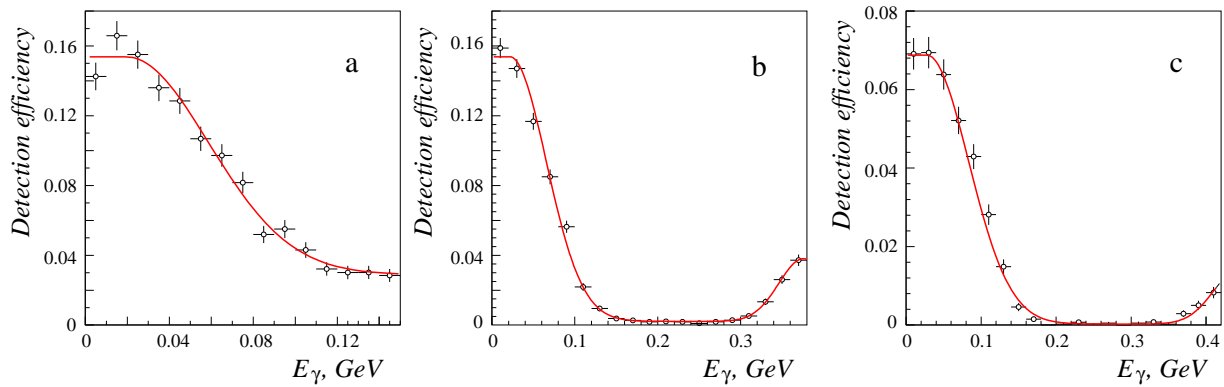}
\caption{ \label{eff}
The dependence of the detection efficiency of the process 
$e^+e^-\to\eta\gamma$, $\eta\to 3\pi^0$ on the energy of the initial state 
radiation photon.
a) $E<0.88$ GeV;
b) $0.88\leq E \leq 1.06$ GeV;
c) $E > 1.06$ GeV.
}
\end{figure}
Detection efficiency for the process under study was determined from 
Monte-Carlo simulation which takes into account the  radiative corrections 
due to photon radiation by initial particles \cite{kuraev}. Angular 
distribution of these photons was modeled according to the work 
\cite{martin}. The detection efficiency was evaluated as a function of two 
parameters: the center-of-mass energy and the energy of the additional 
photon $E_r$. Fig.\ref{eff} shows the dependence of the detection 
efficiency of the process $e^+e^-\to\eta\gamma$, $\eta\to 3\pi^0$ on $E_r$.
When the ``lost energy'' $E_r$ increases, the efficiency falls at first 
and then increases again. Such behavior is due to the presence of two 
different photons in an event: one from the vector meson decay
$V\to\eta\gamma$ and the another from the initial state radiation. 
As the initial state radiation photon reaches its highest allowed energy,
it starts to substitute the decay photon, which becomes soft.
In fact the number of events from this kinematic region is negligible
due to low probability of hard photon emission and very low $e^+e^-\to\eta\gamma$
cross section at threshold.

Detection efficiency values $\epsilon(E,E_r)$ were calculated in 
each energy point for both decay channels of the $\eta$-meson (\ref{etgnc})
and (\ref{etgcc}). The efficiency obtained from the simulation was corrected 
to take into account the difference of the detector response  
simulation from reality. The evaluation of correction 
factors is described in the section~\ref{syser}.

The visible cross section of the process $e^+e^-\to\eta\gamma$ can be written
as
\begin{equation}
\label{viscrs}
\sigma_{vis}(E) = \int\limits_{0}^{x_{max}} \epsilon_r(E,\frac{xE}{2}) F(x,E) \sigma(\sqrt{1-x}E)dx~,
\end{equation}
where $\sigma(E)$ is the Born cross section, which one needs to extract from
the experiment, $F(x,E)$ is a function describing a probability 
distribution of the
energy fraction $x=2E_r/E$~\cite{kuraev} taken away by the additional
photon. Equation (\ref{viscrs}) can be rewritten in the traditional form:
\begin{equation}
\label{viscrs1}
\sigma_{vis}(E) = \epsilon(E)\,\sigma(E)\,(1+\delta(E))~,
\end{equation}
where the parameter $\epsilon(E)$ 
is defined as follows:
\begin{equation}
\epsilon(E) \equiv \epsilon_r(E,0)~,
\end{equation}
and $\delta(E)$ is the radiative correction.

Technically the determination of the Born cross sections is performed as 
follows. With the use of formula (\ref {viscrs}) the energy dependence
of the measured visible cross section is approximated.
For that the Born cross section is parametrized by some theoretical 
model describing the data well. With the use of the 
obtained parameters of the model, the radiative correction is 
determined and then with the formula (\ref {viscrs1}) the experimental
Born cross section $\sigma$ is calculated. To estimate the model dependence
of the radiative correction due to the choice of cross section approximation
function, several models of the Born cross section parametrization are used.

\section{Cross section parameterization}
Energy dependence of the $e^+e^-\to\eta\gamma$ Born cross section
was parameterized according to the vector meson dominance model: 
\begin{equation}
\label{crs1}
\sigma_{\eta\gamma}(E) = \frac{q(E)^3}{E^3} \left| 
\sum\limits_{V=\rho,~\omega,~\phi,\rho^\prime} A_{V}(E)\right|^2,\;\;\;
A_V(E) = \frac{m_V \Gamma_V e^{i\varphi_V}}{D_V(E)} \sqrt{ \frac{m^3_V}
{q(m_V)^3} \sigma_{V\eta\gamma}}~, 
\end{equation}
\[ D_V(E) = m^2_V -E^2 - i E\Gamma_V(E),\;\;\; q(E) = \frac{E}{2} \left( 1 - \frac{m^2_{\eta}}{E^2} \right)~. \]
Here $m_V$ is the resonance mass, $\Gamma_V(E)$ is its full width which 
depends on energy ($\Gamma_V\equiv \Gamma_V(m_V))$,
$\sigma_V =12\pi \BR(V\to e^+e^-)/m^2_V$ and $\sigma_{V\eta\gamma}=
\sigma_V \BR(V\to \eta\gamma)$ are Born cross sections of the 
$e^+e^-\to V$ and $e^+e^-\to V\to\eta\gamma$ processes at $E=m_V$, 
$\BR(V\to e^+e^-)$ and $\BR(V\to \eta\gamma)$ are branching fractions of the 
corresponding decays, $\varphi_V$ are interference phases ($\varphi_{\rho} \equiv 0$).

At approximations of the data, the free parameters were
$\sigma_{\rho\eta\gamma}$,
$\sigma_{\omega\eta\gamma}$,
$\sigma_{\phi\eta\gamma}$ and the phases $\varphi_\omega$, $\varphi_\phi$.
In the energy region above the $\phi$-meson resonance it is necessary to 
take into account contributions from decays of the radial excitations of the
$\rho$, $\omega$ and $\phi$ mesons. As the experimental statistics in this 
energy region is scarce, we restricted ourselves to introduction of just
one additional resonance with the mass $M_{\rho^{\prime}}= 1.465$ GeV and 
width $\Gamma_{\rho^{\prime}}=0.4$ GeV. The cross section at the resonance 
$\sigma_{\rho^{\prime}\eta\gamma}$ and the phase $\varphi_{\rho^{\prime}}$ 
were also free parameters. 

\section{Systematic uncertainties\label{syser}}
The systematic errors on the measured Born cross section include
the contributions of uncertainties in the detection efficiency, 
the luminosity measurement, and systematic errors of the 
background subtraction which were discussed above.

To take into account imperfect  modeling of the detector response, the 
detection efficiency determined from simulation was multiplied by a 
correction factor. It was evaluated using events from 
the energy interval in the vicinity of the
$\phi$-meson resonance where the process (\ref{etg}) can be 
extracted with low background by application of the 
selection criteria less stringent than those described in Section~\ref {evsel}. 
In particular, we can exclude the
condition $\chi^2_{n\gamma}<30 $ and calculate the corresponding correction 
factor (connected with the application of this condition) as follows:
\[
r_{\chi^2_{n\gamma}}=
\frac{N_{data}(\chi^2_{n\gamma}<30)}{N_{MC}(\chi^2_{n\gamma}<30)}
\frac{N_{MC}}{N_{data}}~,
\]
where $N_{data}$, $N_{MC}$ are event numbers in the data and simulation,
respectively, without the cut on $\chi^2_{n\gamma}$,
$N_{data}(\chi^2_{n\gamma}<30)$, $N_{MC}(\chi^2_{n\gamma}<30)$ are
numbers of events with $\chi^2_{n\gamma}<30$. For data 
events the preliminary subtraction of the background from the process  
$e^+e^-\to K_S K_L$ was performed. Analogous corrections were  calculated for 
all other selection conditions used in the analysis.
The overall correction factors for the mode (\ref {etgnc}) were 
$0.990 \pm0.004$ and $0.952 \pm0.004$ for selection criteria used for energies $E<1.06$ GeV and 
$E > 1.06$ GeV, respectively. Since the fraction of events rejected by the
selection cuts is practically independent on energy, the correction found
at $\phi$-meson was applied to all energies. The uncertainty of the correction factor 
was included into the detection efficiency uncertainty.

As it was discussed earlier, additional ``false'' photons appear in the SND
calorimeter because of beam-background pile-up. To take into account
this effect in the Monte-Carlo simulation, actual events recorded with special
random trigger were merged with simulated events in proper proportion.
Unaccounted difference in photon multiplicities between data and simulation can 
lead to an additional uncertainty in the detection efficiency. 
Another effect which is not compensated by the correction factor 
considered above is a possible inaccuracy of the calorimeter response 
modeling near the calorimeter edges. 

To estimate the influence of these effects we analyzed events with 
additional conditions $N_{\gamma}\geq 7$ and $\theta_{\gamma}\geq 36^{\circ}$, 
where $\theta_{\gamma}$ is a polar angle of the detected photon. The 
following ratios have been calculated:
\[ r_{N_{\gamma}} = \frac{N^{data}(N_{\gamma} \geq 7)}{N^{MC}(N_{\gamma}
\geq 7)}~\frac{N^{MC}}{N^{data}} = 0.989 \pm 0.008~, \]
\[ r_{\theta_{\gamma}} = \frac{N^{data}(\theta_{\gamma}>36^{\circ})}
{N^{MC}(\theta_{\gamma}>36^{\circ})}~\frac{N^{MC}}{N^{data}} = 
1.016 \pm 0.010~. \]
Although both ratios are compatible with unity, we use their deviations from 
unity as estimates of the systematic uncertainties connected with 
discrepancies of the photon multiplicity and edge effects modeling.

Thus the total systematic error of the detection efficiency for 
the mode (\ref {etgnc}) is estimated to be 1.9\%.

For the decay mode (\ref {etgcc}), the total correction factor for 
the detection efficiency turned out to be $0.942\pm 0.08$. 
The uncertainty due to the ``edge effect" was taken from the 
analysis of the channel (\ref {etgnc}). The total systematic error for the 
decay mode (\ref {etgcc}) is 1.8\%.

The integrated luminosity was determined by using two QED 
processes $e^+e^-\to e^+e^-$ and $e^+e^-\to\gamma\gamma$ which cross 
sections are known with accuracy better than 1\%. 
The difference between these two luminosity measurements
was used as an estimate of the systematic uncertainty of the luminosity
determination. It is 2\% and practically independent of the beam energy.

The uncertainty in the the radiative correction 
calculation includes the theoretical uncertainty which does not exceed 0.1\% 
\cite {kuraev}, and the model uncertainty, which is also 0.1\%.

\section{Results of the cross section approximation}
Before carrying out approximation of the cross section energy dependence,
the ratio of the $\eta$-meson decay probabilities in two $\eta$ decay modes
was calculated from the corresponding event numbers and detection efficiencies.
It was found that the ratio does not depend on the beam energy. 
Therefore combined approximation of the visible cross sections 
measured in two decay modes was performed. For the decay probabilities 
$\eta\to 3\pi^0$ and $\pi^0\to 2\gamma$ the world-average values~\cite{pdg} were used, 
the ratio of the decay probabilities $\eta\to 3\pi^0$ and $\eta\to\pi^+\pi^-\pi^0$
was a free parameter. As a result of the approximation the following value of 
this ratio has been obtained
\begin{eqnarray}
\frac{\BR(\eta\to 3\pi^0)}{\BR(\eta\to\pi^+\pi^-\pi^0)} 
& = & 1.46 \pm 0.03 \pm 0.09~, \nonumber
\end{eqnarray}
The systematic error includes uncertainties in the detection efficiency, 
the luminosity measurement, and the systematic error in the number
of selected events. Our result is in a good agreement with the world-average
value~\cite{pdg}, $1.44\pm 0.04$. Further approximation was performed with 
this ratio fixed at the world-average value but with its uncertainty taken into account.

\begin{table}
\caption{\label{tabresfit} Results of approximation. The first error is 
statistical, the second systematic.}
\begin{ruledtabular}
\begin{tabular}[t]{cc} 
$\sigma_{\rho\eta\gamma}$                     & $(0.273 \pm 0.029 \pm 0.006)$~nb        \\ 
$\sigma_{\omega\eta\gamma}$                   & $(0.797 \pm 0.079 \pm 0.017)$~nb        \\ 
$\varphi_{\omega}$                            & $(11.3 \pm 8.1 \pm 0.3)^{\circ}$   \\
$\sigma_{\phi\to\eta\gamma}$                  & $(57.14 \pm 0.79 \pm 1.26)$~nb      \\ 
$\varphi_{\phi}$                              & $(170 \pm 12 \pm 4)^{\circ}$ \\ 
$\sigma_{\rho^{\prime}\eta\gamma}$            & $(0.020^{+0.019}_{-0.013} \pm 0.001)$~nb          \\ 
$\varphi_{\rho^{\prime}}$                     & $(61^{+39}_{-20} \pm 2)^{\circ}$  \\ \hline
$\chi^2_{3\pi^0}/ndf$                         & $31/41$          \\ 
$\chi^2_{\pi^+\pi^-\pi^0}/ndf$                & $9.8/19$           \\ 
\end{tabular}
\end{ruledtabular}
\end{table}
Values of the cross sections at the resonance maximums and the interference 
phases obtained  as a result of approximation are presented in the Table 
\ref{tabresfit}.

\begin{figure}
\includegraphics[width=0.8\linewidth]{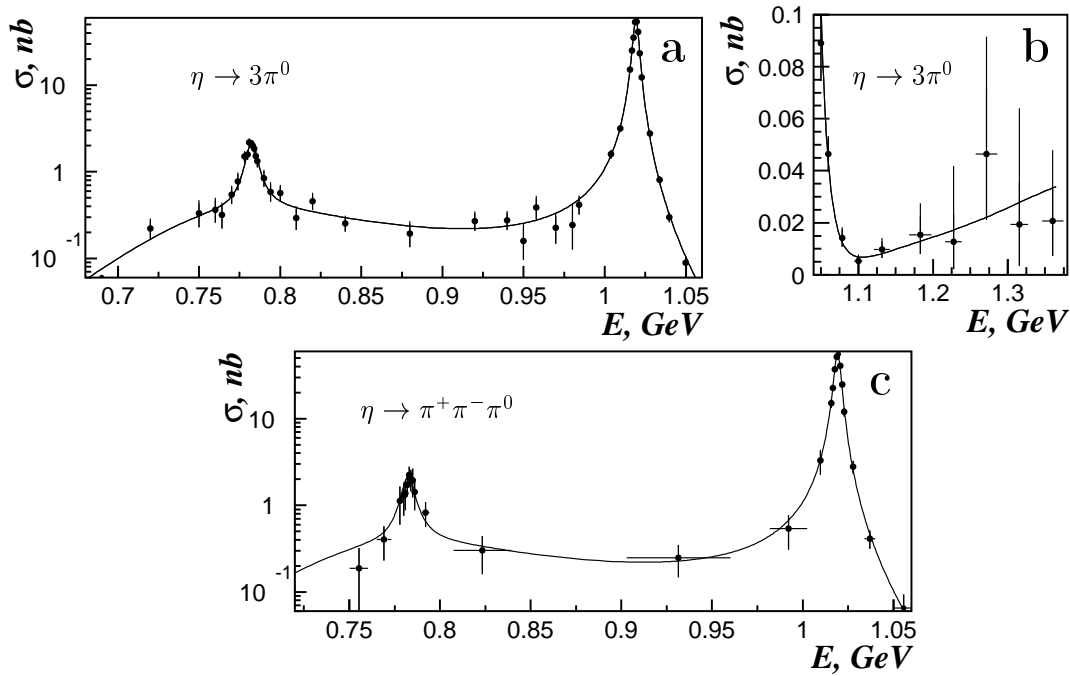}
\caption{ \label{crs}
Cross section of the process $e^+e^-\to\eta\gamma$ 
measured in the $\eta\to 3\pi^0$ decay mode (a,b) and
the $\eta\to\pi^+\pi^-\pi^0$ decay mode (c).}
\end{figure}
The cross section of the $e^+e^-\to\eta\gamma$ process 
in the modes $\eta\to 3\pi^0$ and $\eta\to\pi^+\pi^-\pi^0$ is shown in 
Fig.\ref {crs} and in the Tables \ref{tabcrs1}, \ref{tabcrs2},
and 
\ref {tabcrs3}.
Using the world-average values~\cite{pdg} for the masses and for the decay probabilities 
$\rho,\omega,\phi\to e^+e^-$ we can calculate the decay probabilities into 
the $\eta\gamma$ final state:
\begin{eqnarray}
\BR(\rho\to\eta\gamma)   & =  & (2.40 \pm 0.25 \pm 0.07) \times 10^{-4}, 
\nonumber \\
\BR(\omega\to\eta\gamma) & =  & (4.63 \pm 0.46 \pm 0.13) \times 10^{-4}, 
\nonumber \\
\BR(\phi\to\eta\gamma)   & =  & (1.362 \pm 0.019 \pm 0.035) \times 10^{-2}. 
\nonumber 
\end{eqnarray}

\section{Conclusion}
In the present work the cross section of the process $e^+e^-\to\eta\gamma$ 
was measured in the decay modes $\eta\to 3\pi^0$ and 
$\eta\to\pi^+\pi^-\pi^0$ in the energy range 0.60--1.38 GeV. 
The obtained data agree with the results of previous measurements
~\cite{exd3,exd2,exd23,exd16,exd7,exd17,exd22,exd18}.

\begin{table}
\caption{\label{tabsum} The values of $\rho,\omega,\phi \to \eta\gamma$
branching fractions obtained in this work (second column) and  in the most precise previous
experiments (third column). The first error is 
statistical, the second systematic. The current world-average value are listed in the fourth 
column.}
\begin{ruledtabular}
\begin{tabular}[t]{lccc}
& This work & Previous measurements & PDG~\cite{pdg} \\
$\BR(\rho\to\eta\gamma)\times10^4$   & $2.40 \pm 0.25 \pm 0.07$    & $3.28 \pm 0.37 \pm 0.23$~\cite{exd2}    & $3.0 \pm 0.4$ \\
$\BR(\omega\to\eta\gamma)\times10^4$ & $4.63 \pm 0.46 \pm 0.13$    & $5.10 \pm 0.72 \pm 0.34$~\cite{exd2}    & $4.9 \pm 0.5$ \\
$\BR(\phi\to\eta\gamma)\times10^2$   & $1.362 \pm 0.019 \pm 0.035$ & $1.338 \pm 0.012 \pm 0.52$~\cite{exd17} & $1.295\pm0.025$ \\
\end{tabular}
\end{ruledtabular}
\end{table}
The measured cross section is well described by the vector meson 
dominance model
with contributions from $\rho(770)$, $\omega(783)$, $\phi(1020)$, 
and their excitations, represented in our fit to data by $\rho^{\prime}(1465)$. 
As a result of the data approximation   
we obtained the $\rho,\omega,\phi \to \eta\gamma$ decay probabilities
listed in Table~\ref{tabsum} together with the world-average values
and most precise results of previous measurements. 
The quoted PDG values
include SND measurements in $\eta\to 3\pi^0$ and $\eta\to \pi^+\pi^-\pi^0$ 
modes~\cite{exd7,exd18} based on a part of collected statistics. 
The present work uses full SND statistics 
and supersede the results of the measurements~\cite{exd7,exd18}.

In contrast to previous measurements, the approximation of the 
$e^+e^-\to\eta\gamma$ cross section was performed with free interference 
phases and therefore the obtained values of the decay probabilities do not 
depend on model assumptions on these phases. The following values 
of the phases were obtained:
\begin{eqnarray}
\varphi_{\omega}       & = &  (11.3 \pm 8.1 \pm 0.3)^{\circ}, \nonumber  \\
\varphi_{\phi}         & = &  (170 \pm 12 \pm 4)^{\circ}. \nonumber 
\end{eqnarray}

From the ratio of the $e^+e^-\to\eta\gamma$ cross sections in two
$\eta$ decay modes, the ratio of the $\eta$ meson decay probabilities was 
measured:
\begin{eqnarray}
\frac{\BR(\eta\to 3\pi^0)}{\BR(\eta\to\pi^+\pi^-\pi^0)} & = & 1.46 \pm 0.03 
\pm 0.09~. \nonumber
\end{eqnarray}

The work is supported in part by grants  Sci.School-905.2006.2 and
RFBR
04-02-16181-a,
04-02-16184-a,
05-02-16250-a,
06-02-16292-a,
03-02-16292-a.

\begin{table}
\caption{\label{tabcrs1} Cross section ($\sigma$) of the process 
$e^+e^-\to\eta\gamma$ measured in the decay mode 
$\eta\to 3\pi^0$. $E$ is center-of-mass energy, $\delta E$ is its uncertainty,
$IL$ is integrated luminosity, $N$ is number of selected events with
background subtracted,
$\epsilon_0$ is detection efficiency, $\delta$ is radiative correction.
The first error is statistical, the second systematic.}
\begin{ruledtabular}
\begin{tabular}[t]{ccccccc} 
$E$, MeV & $\delta E$, MeV & $IL$, nb$^{-1}$ & $N$ & $\epsilon_0$ & $\delta+1$ & $\sigma$, nb \\ \hline
  599.94 &  0.16 &   88.22 &    0.0$^{+0.5}_{-0.0}$       & 0.091  & 0.8424 &  $< 0.36~90\% CL$  \\
  629.96 &  0.18 &  117.47 &    0.0$^{+0.5}_{-0.0}$       & 0.129  & 0.8586 &  $< 0.19~90\% CL$  \\
  660.00 &  0.18 &  274.29 &    2.0$^{+1.8}_{-1.0}$       & 0.142  & 0.8671 &  0.059$^{+0.078}_{-0.038}$  \\
  689.98 &  0.20 &  170.46 &    0.0$^{+0.5}_{-0.0}$       & 0.159  & 0.8712 &  $< 0.10~90\% CL$ \\
  720.00 &  0.21 &  575.44 &   17.0$^{+4.5}_{-3.8}$       & 0.152  & 0.8754 &  0.22$^{+0.07}_{-0.05}$  \\
  749.98 &  0.23 &  225.20 &   10.0$^{+3.5}_{-2.8}$       & 0.152  & 0.8837 &  0.33$^{+0.14}_{-0.10}$  \\
  760.00 &  0.24 &  246.09 &   12.0$^{+3.8}_{-3.1}$       & 0.152  & 0.8848 &  0.36$^{+0.14}_{-0.10}$  \\
  764.00 &  0.26 &  258.70 &   11.0$^{+3.7}_{-3.0}$       & 0.152  & 0.8825 &  0.32$^{+0.13}_{-0.09}$  \\
  770.00 &  0.25 &  291.87 &   21.0$^{+4.9}_{-4.3}$       & 0.152  & 0.8689 &  0.54$^{+0.15}_{-0.12}$  \\
  773.98 &  0.25 &  220.07 &   22.0$^{+5.0}_{-4.4}$       & 0.152  & 0.8452 &  0.78$^{+0.20}_{-0.17}$  \\
  778.00 &  0.30 &  259.50 &   48.0 $\pm$  6.9            & 0.152  & 0.8068 &  1.50  $\pm$ 0.22  \\
  780.00 &  0.25 &  315.38 &   60.0 $\pm$  7.7            & 0.152  & 0.7922 &  1.58  $\pm$ 0.21  \\
  780.98 &  0.24 &  346.63 &   91.0 $\pm$  9.5            & 0.152  & 0.7913 &  2.18  $\pm$ 0.23  \\
  782.00 &  0.24 &  664.82 &  171.0 $\pm$ 13.1            & 0.152  & 0.7976 &  2.13  $\pm$ 0.17  \\
  783.00 &  0.24 &  483.13 &  119.0 $\pm$ 10.9            & 0.152  & 0.8118 &  2.00  $\pm$ 0.19  \\
  783.98 &  0.24 &  356.57 &   83.0 $\pm$  9.1            & 0.152  & 0.8324 &  1.84  $\pm$ 0.21  \\
  785.02 &  0.24 &  221.39 &   44.0 $\pm$  6.6            & 0.152  & 0.8585 &  1.52  $\pm$ 0.23  \\
  785.98 &  0.24 &  270.02 &   48.0 $\pm$  6.9            & 0.152  & 0.8835 &  1.32  $\pm$ 0.19  \\
  789.96 &  0.25 &  193.41 &   24.0$^{+5.2}_{-4.6}$       & 0.152  & 0.9651 &  0.85$^{+0.21}_{-0.17}$  \\
  793.96 &  0.25 &  211.95 &   19.0$^{+4.7}_{-4.0}$       & 0.152  & 1.004  &  0.59$^{+0.17}_{-0.13}$  \\
  800.00 &  0.25 &  284.30 &   25.0$^{+5.3}_{-4.7}$       & 0.152  & 1.021  &  0.57$^{+0.14}_{-0.11}$  \\
  809.98 &  0.26 &  286.94 &   13.0$^{+3.9}_{-3.3}$       & 0.152  & 1.021  &  0.29$^{+0.11}_{-0.08}$  \\
  819.94 &  0.26 &  321.32 &   23.0$^{+5.1}_{-4.5}$       & 0.155  & 1.010  &  0.46$^{+0.12}_{-0.09}$  \\
  840.00 &  0.28 &  692.59 &   27.0$^{+5.5}_{-4.9}$       & 0.155  & 0.9923 &  0.25$^{+0.06}_{-0.05}$  \\
  879.92 &  0.31 &  384.26 &   11.0$^{+3.7}_{-3.0}$       & 0.155  & 0.9594 &  0.19$^{+0.08}_{-0.06}$  \\
  919.86 &  0.35 &  487.38 &   19.0$^{+4.7}_{-4.0}$       & 0.155  & 0.9312 &  0.27$^{+0.08}_{-0.06}$  \\
  939.88 &  0.34 &  488.74 &   19.0$^{+4.7}_{-4.0}$       & 0.155  & 0.9155 &  0.27$^{+0.08}_{-0.06}$  \\
  949.78 &  0.32 &  268.39 &    6.0$^{+2.8}_{-2.1}$       & 0.155  & 0.9065 &  0.16$^{+0.10}_{-0.06}$  \\
  957.80 &  0.32 &  241.85 &   13.0$^{+3.9}_{-3.3}$       & 0.155  & 0.8981 &  0.39$^{+0.14}_{-0.11}$  \\ 
  969.80 &  0.34 &  258.55 &    8.0$^{+3.2}_{-2.5}$       & 0.155  & 0.8833 &  0.23$^{+0.11}_{-0.08}$  \\
  980.00 &  0.21 &  124.62 &    4.0$^{+2.3}_{-1.7}$       & 0.152  & 0.8703 &  0.24$^{+0.19}_{-0.12}$  \\
  984.10 &  0.37 &  348.02 &   19.0$^{+4.7}_{-4.0}$       & 0.152  & 0.8624 &  0.42$^{+0.12}_{-0.09}$  \\
 1003.82 &  0.38 &  365.59 &   72.0 $\pm$  8.5 $\pm$  0.3 & 0.152  & 0.8061 &  1.60  $\pm$ 0.19  \\
 1009.68 &  0.39 &  299.53 &  113.0 $\pm$ 10.6 $\pm$  0.5 & 0.152  & 0.7771 &  3.18  $\pm$ 0.31  \\
 1015.64 &  0.39 &  344.73 &  588.3 $\pm$ 24.3 $\pm$  2.8 & 0.152  & 0.7324 & 15.10  $\pm$ 0.89  \\
 1016.70 &  0.38 &  601.93 & 1680.7 $\pm$ 41.0 $\pm$  8.1 & 0.152  & 0.7233 & 24.99  $\pm$ 1.26  \\
 1017.66 &  0.38 &  937.42 & 3659.2 $\pm$ 60.5 $\pm$ 17.6 & 0.152  & 0.7182 & 35.44  $\pm$ 1.79  \\
 1018.64 &  0.39 &  984.89 & 5691.5 $\pm$ 75.4 $\pm$ 27.3 & 0.152  & 0.7238 & 53.53  $\pm$ 1.73  \\
 1019.62 &  0.42 & 1060.53 & 6365.2 $\pm$ 79.8 $\pm$ 30.6 & 0.152  & 0.7544 & 54.02  $\pm$ 1.29  \\
 1020.58 &  0.40 &  628.02 & 3213.9 $\pm$ 56.7 $\pm$ 15.4 & 0.152  & 0.8136 & 41.31  $\pm$ 1.70  \\
 1021.64 &  0.41 &  325.47 & 1056.3 $\pm$ 32.5 $\pm$  5.1 & 0.152  & 0.9012 & 23.32  $\pm$ 1.22  \\
 1022.78 &  0.39 &  353.27 &  672.4 $\pm$ 25.9 $\pm$  3.2 & 0.152  & 1.009  & 12.24  $\pm$ 0.70  \\
 1027.76 &  0.40 &  362.76 &  241.2 $\pm$ 15.5 $\pm$  1.2 & 0.152  & 1.577  &  2.76  $\pm$ 0.19  \\
 1033.70 &  0.39 &  327.43 &   98.8 $\pm$  9.9 $\pm$  0.5 & 0.152  & 2.468  &  0.80  $\pm$ 0.08  \\
 1039.68 &  0.39 &  389.43 &   65.0 $\pm$  8.1 $\pm$  0.3 & 0.152  & 3.669  &  0.30  $\pm$ 0.04  \\
 1049.76 &  0.39 &  441.28 &   39.3 $\pm$  6.3 $\pm$  0.2 & 0.152  & 6.584  &  0.089 $\pm$ 0.014 \\
 1059.76 &  0.44 &  637.23 &   48.3 $\pm$  7.0 $\pm$  0.2 & 0.152  & 10.74  &  0.046 $\pm$ 0.007 \\
\end{tabular}
\end{ruledtabular}
\end{table}
\begin{table}
\caption{\label{tabcrs2} Cross section ($\sigma$) of the process 
$e^+e^-\to\eta\gamma$ measured in the decay mode 
$\eta\to 3\pi^0$. $E$ is center-of-mass energy, $\delta E$ is its uncertainty,
$IL$ is integrated luminosity, $N$ is number of selected events with
background subtracted,
$\epsilon_0$ is detection efficiency, $\delta$ is radiative correction.
The first error is statistical, the second systematic.}
\begin{ruledtabular}
\begin{tabular}[t]{ccccccc} 
$E$, MeV & $\delta E$, MeV & $IL$, nb$^{-1}$ & $N$ & $\epsilon_0$ & $\delta+1$ & $\sigma$, nb \\ \hline
 1078.54 &  3.55 &  650.05 &   18.0$^{+4.6}_{-3.9}$       & 0.0789 & 24.65  &  0.014$^{+0.004}_{-0.003}$  \\
 1099.92 &  5.45 &  605.60 &    9.0$^{+3.3}_{-2.7}$       & 0.0769 & 35.73  &  0.005$^{+0.002}_{-0.002}$  \\
 1131.58 & 10.49 &  749.68 &    9.0$^{+3.3}_{-2.7}$       & 0.0739 & 16.39  &  0.010$^{+0.004}_{-0.003}$  \\
 1182.96 & 15.08 & 1292.02 &    4.0$^{+2.3}_{-1.7}$       & 0.0692 & 2.816  &  0.015$^{+0.012}_{-0.007}$  \\
 1227.34 & 11.14 &  959.07 &    1.0$^{+1.4}_{-0.5}$       & 0.0651 & 1.254  &  0.013$^{+0.029}_{-0.010}$  \\
 1271.68 & 14.33 & 1061.97 &    3.0$^{+2.1}_{-1.4}$       & 0.0610 & 0.9952 &  0.046$^{+0.045}_{-0.025}$  \\
 1315.44 & 11.72 &  954.60 &    1.0$^{+1.4}_{-0.5}$       & 0.0570 & 0.9465 &  0.019$^{+0.045}_{-0.016}$  \\
 1360.44 & 14.39 & 1958.92 &    2.0$^{+1.8}_{-1.0}$       & 0.0529 & 0.9335 &  0.021$^{+0.027}_{-0.013}$  \\
\end{tabular}
\end{ruledtabular}
\end{table}
\begin{table}
\caption{\label{tabcrs3} 
Cross section ($\sigma$) of the process $e^+e^-\to\eta\gamma$  measured in 
the decay mode $\eta\to\pi^+\pi^-\pi^0$.
$E$ is center-of-mass energy, $\delta E$ is its uncertainty,
$IL$ is integrated luminosity, $N$ is number of selected events with
background subtracted,
$\epsilon_0$ is detection efficiency, $\delta$ is radiative correction.
The first error is statistical, the second systematic.}
\begin{ruledtabular}
\begin{tabular}[t]{ccccccc} 
$E$, MeV & $\delta E$, MeV & $IL$, nb$^{-1}$ & $N$ & $\epsilon_0$ & $\delta+1$ & $\sigma$, nb \\ \hline
 755.26 & 5.00 &  492.5 &    5.7 $\pm$  4.0 $\pm$  1.0 & 0.0695 & 0.8808 &   0.19  $\pm$ 0.13  \\
 769.12 & 3.98 &  809.3 &   20.7 $\pm$  8.8 $\pm$  1.7 & 0.0695 & 0.8683 &   0.40  $\pm$ 0.17  \\
 777.96 & 0.24 &  264.6 &   16.6 $\pm$  7.7 $\pm$  0.3 & 0.0695 & 0.8055 &   1.12  $\pm$ 0.52  \\
 779.98 & 0.28 &  338.5 &   22.1 $\pm$  9.1 $\pm$  0.4 & 0.0640 & 0.7918 &   1.28  $\pm$ 0.53  \\
 780.98 & 0.24 &  365.3 &   25.6 $\pm$  9.4 $\pm$  0.5 & 0.0640 & 0.7909 &   1.38  $\pm$ 0.51  \\
 782.00 & 0.24 &  697.3 &   61.1 $\pm$ 16.4 $\pm$  1.2 & 0.0640 & 0.7972 &   1.71  $\pm$ 0.46  \\
 783.00 & 0.24 &  506.3 &   59.1 $\pm$ 14.3 $\pm$  1.2 & 0.0640 & 0.8114 &   2.23  $\pm$ 0.54  \\
 783.98 & 0.24 &  374.3 &   40.0 $\pm$ 11.1 $\pm$  0.8 & 0.0640 & 0.8319 &   1.98  $\pm$ 0.55  \\
 785.02 & 0.24 &  232.0 &   24.8 $\pm$  9.0 $\pm$  0.5 & 0.0640 & 0.8578 &   1.92  $\pm$ 0.70  \\
 785.98 & 0.24 &  280.7 &   22.6 $\pm$  8.7 $\pm$  0.5 & 0.0640 & 0.8826 &   1.40  $\pm$ 0.54  \\
 792.06 & 2.02 &  421.9 &   22.7 $\pm$  7.3 $\pm$  0.5 & 0.0640 & 0.9875 &   0.81  $\pm$ 0.26  \\
 823.36 & 15.9 & 1658.6 &   33.4 $\pm$ 15.6 $\pm$  0.7 & 0.0640 & 1.005  &   0.30  $\pm$ 0.14  \\
 931.52 & 28.7 & 2273.1 &   36.7 $\pm$ 14.9 $\pm$  0.7 & 0.0640 & 0.9319 &   0.25  $\pm$ 0.10  \\
 992.14 & 10.4 &  907.2 &   25.8 $\pm$ 10.9 $\pm$  0.5 & 0.0500 & 0.8480 &   0.54  $\pm$ 0.23  \\
1009.68 & 0.39 &  325.5 &   42.0 $\pm$ 13.2 $\pm$  0.8 & 0.0500 & 0.7778 &   3.30  $\pm$ 1.04  \\
1015.64 & 0.39 &  374.7 &  209.9 $\pm$ 21.9 $\pm$  4.2 & 0.0500 & 0.7325 &  15.07  $\pm$ 1.74  \\
1016.70 & 0.38 &  660.0 &  550.7 $\pm$ 35.9 $\pm$ 11.0 & 0.0500 & 0.7234 &  22.70  $\pm$ 1.95  \\
1017.66 & 0.38 & 1028.1 & 1383.3 $\pm$ 54.1 $\pm$ 27.7 & 0.0500 & 0.7182 &  37.14  $\pm$ 2.42  \\
1018.64 & 0.39 & 1080.9 & 1989.2 $\pm$ 65.4 $\pm$ 39.8 & 0.0500 & 0.7238 &  51.83  $\pm$ 2.68  \\
1019.62 & 0.42 & 1159.4 & 2357.6 $\pm$ 72.6 $\pm$ 47.2 & 0.0500 & 0.7544 &  55.63  $\pm$ 2.46  \\
1020.58 & 0.40 &  687.1 & 1144.3 $\pm$ 49.3 $\pm$ 22.9 & 0.0500 & 0.8136 &  40.87  $\pm$ 2.54  \\
1021.64 & 0.41 &  356.3 &  407.3 $\pm$ 29.4 $\pm$  8.1 & 0.0500 & 0.9012 &  24.97  $\pm$ 2.13  \\
1022.78 & 0.39 &  386.9 &  237.9 $\pm$ 24.3 $\pm$  4.8 & 0.0500 & 1.009  &  12.01  $\pm$ 1.37  \\
1027.76 & 0.40 &  392.2 &   86.9 $\pm$ 14.5 $\pm$  1.7 & 0.0500 & 1.577  &   2.79  $\pm$ 0.47  \\
1036.96 & 3.00 &  777.5 &   52.1 $\pm$ 12.2 $\pm$  1.0 & 0.0500 & 3.078  &   0.41  $\pm$ 0.10  \\
1055.64 & 4.94 & 1152.4 &   35.7 $\pm$ 15.2 $\pm$  0.7 & 0.0500 & 9.139  &   0.064 $\pm$ 0.027 \\
\end{tabular}
\end{ruledtabular}
\end{table}


\begin{thebibliography}{99}
%
\bibitem{pdg}
Review of Particle Physics, S.~Eidelman {\it et al.},
Phys.\ Lett.\ B {\bf 592}, 1 (2004).

\bibitem{link1}
P.~O'Donnel, Rev. Mod. Phys. {\bf 53}, 673 (1981).

\bibitem{link2}
G.~Morpurgo, Phys. Rev. D {\bf 42}, 1497 (1990).

\bibitem{link3}
M.~Benayoun {\it et al.}, 
Phys. Rev. D {\bf 59}, 114027 (1999).

\bibitem{QCD1}
M.~Napsuciale, S.~Rodriguez and E.~Alvarado-Anell,
Phys. Rev. D {\bf 67}, 036007  (2003).

\bibitem{QCD2}
C.~Aydin and A.~H.~Yilmaz,
Acta Phys. Polon. B {\bf 34}, 4145 (2003).

\bibitem{QCD3}
R.~Bonnaz, B.~Silvestre-Brac and C.~Gignoux,
Eur. Phys. J. A {\bf 13}, 363 (2002).

\bibitem{mixing1}
R.~Escribano, arXiv:hep-ph/0510206;
R.~Escribano and J.~M.~Frere, JHEP {\bf 0506},  029 (2005);
T.~Feldmann, P.~Kroll and B.~Stech,
Phys.\ Rev.\ D {\bf 58}, 114006 (1998);
T.~Feldmann and P.~Kroll,
Phys.\ Scripta {\bf T99}, 13 (2002).

\bibitem{mixing2}
F.~Kleefeld, arXiv:nucl-th/0510017.

\bibitem{exd3}
R.R.Akhmetshin {\it et al}.,
Phys. Lett. B {\bf 460}, 242 (1999).

\bibitem{exd2}
R.R.Akhmetshin {\it et al}.,
Phys. Lett. B {\bf 509}, 217 (2001).
%
\bibitem{exd23}
R.R.Akhmetshin {\it et al}.,
Phys. Lett. B {\bf 605}, 26 (2005).
%
\bibitem{exd16}
M.N.Achasov {\it et al}.,
JETP Lett. {\bf 68}, 573 (1998).

\bibitem{exd7}
M.N.Achasov {\it et al}.,
JETP Lett. {\bf 72}, 282 (2000).

\bibitem{exd17}
M.N.Achasov {\it et al}.,
Eur. Phys. J. C {\bf 12}, 25 (2000).

\bibitem{exd22}
M.N.Achasov {\it et al}.,
Nucl. Phys. A {\bf 675}, 213 (2000).

\bibitem{exd18}
M.N.Achasov {\it et al}.,
J. Exp. Theor. Phys. {\bf 90}, 17 (2000).

\bibitem{SND}
M.N.Achasov {\it et al}.,
Nucl. Instrum. Methods Phys. Res. A {\bf 449}, 125 (2000).

\bibitem{oppr}
R.R.Akhmetshin {\it et al}.,
Phys.\ Lett.\ B {\bf 489}, 125 (2000)

\bibitem{nfun} The ``Novosibirsk'' function is defined as
$f(x)=A\exp (-0.5{\ln^2[1+\Lambda \tau (x-x_0)]/\tau^2+\tau^2})$,
where $\Lambda=\sinh (\tau\sqrt{\ln{4}})/(\sigma\tau\ln{4})$, $x_0$ is
the peak position, $\sigma$ is the width of the distribution, $\tau$ is 
the asymmetry parameter.
%
\bibitem{kuraev}
E.A.~Kuraev and V.S.~Fadin,
Sov. J. Nucl. Phys. {\bf 41}, 466 (1985).

\bibitem{martin}
G.~Bonneau and F~.Martin,
Nucl. Phys. B {\bf 27} 381 (1971).

\end{thebibliography}
\end{document}